\documentclass[twocolumn]{aastex62}

\newcommand{{\maxi}}{{\it MAXI}}
\newcommand{{\swift}}{{\it Swift}}
\newcommand{\srcname}{{MAXI J1820$+$070 }}

\newcommand{\nh}{\ensuremath{N_\mathrm{H}}}

\newcommand{\Tin}{T_\mathrm{in}}

\graphicspath{{./}{figures/}}

\accepted{October 17, 2018}

\shorttitle{Multi-wavelength Monitoring of MAXI J1820$+$070}
\shortauthors{Shidatsu et al.}

\begin{document}

\title{X-ray, Optical, and Near-infrared Monitoring of the New X-ray Transient MAXI J1820$+$070 in the Low/hard State}

\correspondingauthor{Megumi Shidatsu}
\email{shidatsu.megumi.wr@ehime-u.ac.jp}

\author[0000-0001-8195-6546]{Megumi Shidatsu}
\affil{Department of Physics, Ehime University, 
2-5, Bunkyocho, Matsuyama, Ehime 790-8577, Japan}
\author{Satoshi Nakahira}
\affil{High Energy Astrophysics Laboratory, RIKEN, 2-1, Hirosawa, Wako, Saitama 351-0198, Japan}
\author{Satoshi Yamada}
\affil{Department of Astronomy, Kyoto University, Kitashirakawa-Oiwake-cho, Sakyo-ku, Kyoto, Kyoto 606-8502, Japan}
\author[0000-0002-6808-2052]{Taiki Kawamuro}
\affil{National Astronomical Observatory of Japan, Osawa, Mitaka, Tokyo 181-8588, Japan}
\author[0000-0001-7821-6715]{Yoshihiro Ueda}
\affil{Department of Astronomy, Kyoto University, Kitashirakawa-Oiwake-cho, Sakyo-ku, Kyoto, Kyoto 606-8502, Japan}
\author{Hitoshi Negoro}
\affil{Department of Physics, Nihon University, 1-8-14 Kanda-Surugadai, Chiyoda-ku, Tokyo 101-8308, Japan}
\author{Katsuhiro L. Murata}
\affil{Department of Physics, Tokyo Institute of Technology, 2-12-1 Ookayama, Meguro-ku, Tokyo 152-8551, Japan}
\author{Ryosuke Itoh}
\affil{Department of Physics, Tokyo Institute of Technology, 2-12-1 Ookayama, Meguro-ku, Tokyo 152-8551, Japan}
\author[0000-0001-6584-6945]{Yutaro Tachibana}
\affil{Department of Physics, Tokyo Institute of Technology, 2-12-1 Ookayama, Meguro-ku, Tokyo 152-8551, Japan}
\author{Ryo Adachi}
\affil{Department of Physics, Tokyo Institute of Technology, 2-12-1 Ookayama, Meguro-ku, Tokyo 152-8551, Japan}
\author{Yoichi Yatsu}
\affil{Department of Physics, Tokyo Institute of Technology, 2-12-1 Ookayama, Meguro-ku, Tokyo 152-8551, Japan}
\author[0000-0001-9656-0261]{Nobuyuki Kawai}
\affil{Department of Physics, Tokyo Institute of Technology, 2-12-1 Ookayama, Meguro-ku, Tokyo 152-8551, Japan}
\author{Hidekazu Hanayama}
\affil{Ishigakijima Astronomical Observatory, National Astronomical Observatory of Japan, National Institutes of Natural Sciences, 1024-1 Arakawa, Ishigaki, Okinawa, 907-0024, Japan}
\author{Takashi Horiuchi}
\affil{Ishigakijima Astronomical Observatory, National Astronomical Observatory of Japan, National Institutes of Natural Sciences, 1024-1 Arakawa, Ishigaki, Okinawa, 907-0024, Japan}
\author{Hiroshi Akitaya}
\affil{Graduate School of Science and Engineering, Saitama University, Simo-okubo 135, Sakura-ku, Saitama 338-8570, Japan}
\author{Tomoki Saito}
\affil{Nishi-Harima Astronomical Observatory, Center for Astronomy, University of Hyogo, 407-2, Nishigaichi, Sayo, Hyogo, 679-5313, Japan}
\author{Masaki Takayama}
\affil{Nishi-Harima Astronomical Observatory, Center for Astronomy, University of Hyogo, 407-2, Nishigaichi, Sayo, Hyogo, 679-5313, Japan}
\author{Tomohito Ohshima}
\affil{Nishi-Harima Astronomical Observatory, Center for Astronomy, University of Hyogo, 407-2, Nishigaichi, Sayo, Hyogo, 679-5313, Japan}
\author{Noriyuki Katoh}
\affil{Nishi-Harima Astronomical Observatory, Center for Astronomy, University of Hyogo, 407-2, Nishigaichi, Sayo, Hyogo, 679-5313, Japan}
\author{Jun Takahashi}
\affil{Nishi-Harima Astronomical Observatory, Center for Astronomy, University of Hyogo, 407-2, Nishigaichi, Sayo, Hyogo, 679-5313, Japan}
\author{Takahiro Nagayama}
\affil{Graduate School of Science and Engineering, Kagoshima University, 1-21-35 Korimoto, Kagoshima 890-0065, Japan}
\author{Masayuki Yamanaka}
\affil{Hiroshima Astrophysical Science Center, Hiroshima University, Higashi-Hiroshima, Hiroshima 739-8526, Japan}
\author{Miho Kawabata}
\affil{Department of Physical Science, Hiroshima University, Kagamiyama
1-3-1, Higashi-Hiroshima 739-8526, Japan}
\author{Tatsuya Nakaoka}
\affil{Department of Physical Science, Hiroshima University, Kagamiyama
1-3-1, Higashi-Hiroshima 739-8526, Japan}
\author{Seiko Takagi}
\affil{Department of Cosmosciences, Hokkaido University, Kita 10 Nishi 8, Kita-ku, Sapporo, Hokkaido 060-0810, Japan}
\author{Tomoki Morokuma}
\affil{Institute of Astronomy, Graduate School of Science, The University of Tokyo, 2-21-1, Osawa, Mitaka, Tokyo 181-0015, Japan}
\author{Kumiko Morihana}
\affil{Graduate School of Science, Nagoya University, Furo-cho, Chikusa-ku, Nagoya 464-8602, Japan}
\author{Hiroyuki Maehara}
\affil{Okayama Observatory, Graduate School of Science, Kyoto University, 3037-5 Honjo, Kamogata, Asakuchi, Okayama, 719-0232, Japan}
\author{Kazuhiro Sekiguchi}
\affil{National Astronomical Observatory of Japan, National Institutes of Natural Sciences, 2-21-1, Osawa, Mitaka, Tokyo 181-8588, Japan}

\begin{abstract}

We report X-ray, optical, and near-infrared monitoring of 
the new X-ray transient MAXI J1820$+$070 discovered with 
MAXI on 2018 March 11. Its X-ray intensity reached  
$\sim 2$ Crab in 2--20 keV at the end of March, and then 
gradually decreased until the middle of June. 
In this period, the X-ray spectrum was described 
by Comptonization of the disk emission, 
with a photon index of $\sim$1.5 and an electron 
temperature of $\sim$50 keV, which is consistent with 
a black hole X-ray binary in the low/hard state. 
The electron temperature and the photon index were 
slightly decreased and increased with increasing flux, 
respectively. The source showed significant X-ray flux 
variation on timescales of seconds. This short-term 
variation was found to be associated with changes in 
the spectral shape, and the photon index became slightly 
harder at higher fluxes. This suggests that the variation 
was produced by a change in the properties of the hot 
electron cloud responsible for the strong Comptonization. 
Modeling a multi-wavelength SED around the X-ray 
flux peak at the end of March, covering the near-infrared 
to X-ray bands, we found that the optical and near-infrared 
fluxes were 
likely contributed substantially by the jet emission.
Before this outburst, the source was never detected 
in the X-ray band with MAXI (with a 3$\sigma$ upper 
limit of $\sim$0.2 mCrab in 4--10 keV, obtained from 
the 7-year data in 2009--2016), whereas weak optical and 
infrared activity was found at their flux levels 
$\sim$3 orders of magnitude lower than the peak 
fluxes in the outburst. 

\end{abstract}


\section{Introduction} \label{sec:intro}

Transient Galactic black hole binaries (BHBs) 
provides opportunities to study the evolution of 
black hole accretion flows over a wide range of 
mass accretion rates \citep[e.g.,][for reviews]{mcc06,don07}. 
They are usually too faint 
to detect in the X-ray band, but suddenly increase 
their X-ray luminosity by orders of magnitude 
on timescales of days to weeks. At low luminosities, 
they stay in the so-called low/hard state and 
show a power-law shaped hard spectrum, often with 
an exponential cutoff at $\sim$100 keV. This spectral 
profile is often interpreted as thermal Comptonization of 
the soft X-ray photons from the truncated standard 
disk, in a hot electron cloud developed somewhere 
around the disk. However, the geometry of the 
Comptonized region is not yet clear. Moreover, 
energetic electrons and the synchrotron emission 
produced in jets may contribute to the Comptonized 
component, but to what extent they do is still in debate. 
Multi-wavelength observations are important to 
tackle these questions, because the main part 
of the jet synchrotron emission is normally 
located in the radio to optical band.

\begin{figure*}[th!]
\plotone{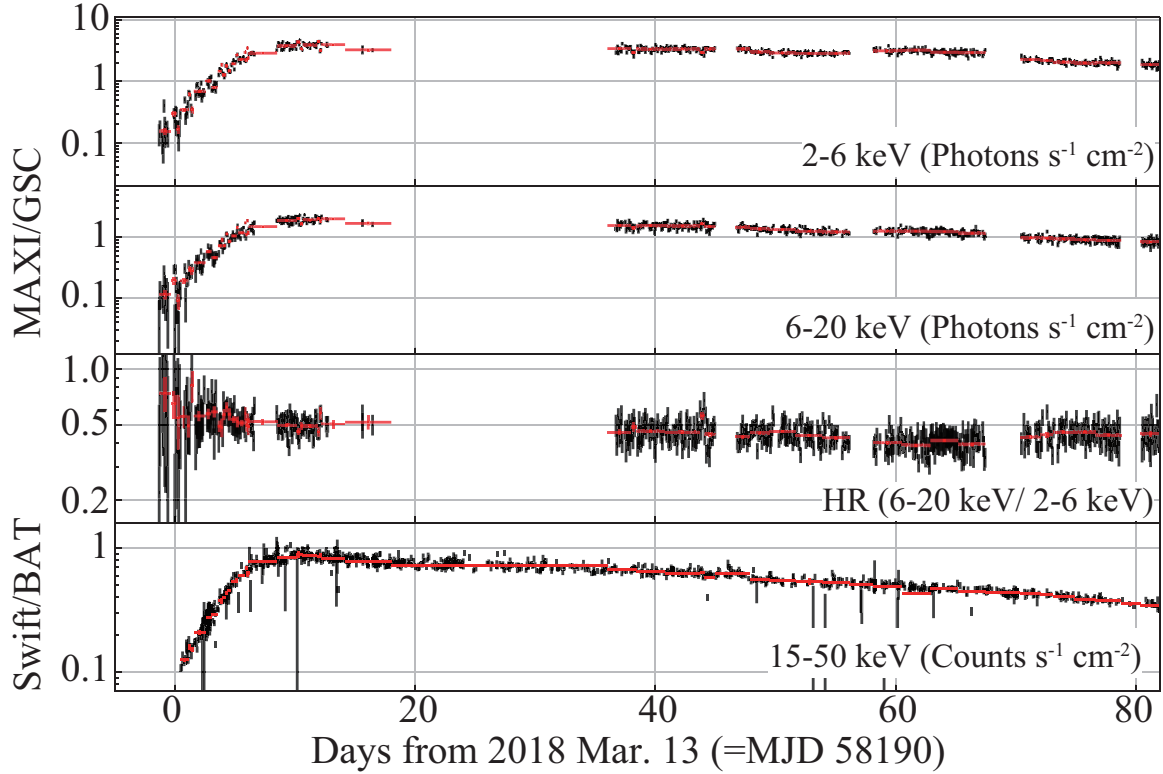}
\caption{\maxi/GSC light curves of MAXI J1820$+$070 in 2--6 keV, 
6--20 keV, their hardness ratio, and \swift/BAT light curve 
in 15--50 keV, from top to bottom. The black points present 
the data with orbital time bins ($\sim$92 min) and the red 
points shows binned data. The error bars represent 
1$\sigma$ statistical errors. \label{fig:LC_longterm}}\end{figure*}

MAXI J1820$+$070 was discovered with the MAXI 
\citep{mat09}/Gas Slit Camera \citep[GSC;][]{mih11}. 
The MAXI Nova Search System \citep{neg16} first 
triggered the source at 2018 March 11 UT 12:50 
\citep{kaw18a}. Soon after the discovery, the source 
was associated with the optical variable source, 
ASSASN-2018ey \citep{den18}. The position of the X-ray source 
was precisely determined in follow-up observations 
with Swift, as $(\alpha^{2000},$ $\delta^{2000}) = 
(18^{\mathrm h}20^{\mathrm m}21^{\mathrm s}.88, 
+07^\circ11'08.''3)$, which was 
consistent with the position of ASSASN-2018ey. 

The relatively small interstellar 
absorption/extinction, with a hydrogen 
column density of $\sim 10^{21}$ cm$^{-2}$, 
and the high flux, exceeding 1 Crab 
in 2--20 keV at the peak, have motivated 
extensive multi-wavelength follow-up 
observations of MAXI J1820$+$070 during 
the outburst \citep{ken18a,ken18b,ken18c,bag18,bri18,
lit18,utt18,bah18,gar18,sak18,del18,pai18,gan18a,tru18,
tet18,ber18,cas18,man18a,man18b,boz18,flo18,mun18,
mer18,kuu18,rus18}. 
Until the middle of June, the source always showed a power-law 
shaped X-ray spectrum with a photon index of $\sim$1.5, 
consistent with BHB spectra in the low/hard state. 
The source was found to show strong optical and 
X-ray short-term variability on time scales 
of less than 1 s \citep{gan18a, sak18}, and in both 
X-ray and optical bands, low-frequency quasi periodic 
oscillations (QPOs) were detected 
at 10--50 mHz \citep{gan18a, mer18, yu18}. 
Radio counterpart was also detected 
\citep{bri18,tru18,tet18}, suggesting the presence of jets.
After the X-ray flux decay until 
the middle of June, the source started to increase 
its X-ray flux again. Then, it showed an X-ray spectral 
softening in July, at the flux comparable to the 
first peak \citep{hom18}.

In this article, we 
investigate the nature of MAXI J1820$+$070, 
mainly focusing on the 
period before the X-ray re-brightening, using X-ray data 
obtained from monitoring observations with MAXI, 
Swift, and optical and near-infrared (IR) data from  
ground-based telescopes participating in the collaboration 
of Optical and Infrared Synergetic Telescopes for Education 
and Research (OISTER) in Japan. Throughout this work, 
we utilized HEASoft version 6.23 for the X-ray data reduction, 
and XSPEC version 12.10.0 with the solar abundance table 
given by \citet{wil00} for the spectral analysis.
Errors represent the 90\% confidence ranges for one parameter, 
unless otherwise stated. 

\section{X-ray Data}
\subsection{Observations and data reduction}
\subsubsection{MAXI data} \label{sec:gsc_reduction}

We reduced the $\maxi$/GSC event data with the 
processed version 1.3.6.6, 
through the $\maxi$ analysis tools implemented 
in ``MAXI/GSC on-demand web interface'' 
\footnote{http://maxi.riken.jp/mxondem} \citep{nak13}.  
The source events were extracted from a circular region with 
a radius of 2$^\circ$.0, centered at the target position. 
Background events were collected from the source-free 
region within 3$^\circ$.0 from the source position, 
determined by excluding the source region 
and 2$^\circ$.0 from nearby bright sources.

Figure~\ref{fig:LC_longterm} shows the \maxi/GSC 
light curves of \srcname in 2--6 keV and 6-20 keV and 
their hardness ratios (HRs), together with the Swift/BAT 
light curve in 15--50 keV, downloaded from the ``BAT Transient Monitor'' 
website \citep{kri13}\footnote{\url{http://swift.gsfc.nasa.gov/docs/swift/results/transients}}, with a time bin size of their orbital periods 
($\sim$92 min; black points in Fig.~\ref{fig:LC_longterm}).  
The soft and hard X-ray fluxes rapidly increased in the 
initial phase of the outburst, and around March 20, it 
reached its peak level of $\sim$5 photons s$^{-1}$ cm$^{-2}$ in 2--20 keV, 
corresponding to $\sim$2 Crab. The source started 
dimming in early April, and from the middle of June, 
it increased its flux again. The HR was almost constant 
before the re-brightening.
To reduce the statistical errors, we binned 1--40 adjacent 
data points that have similar flux levels, as shown in red in 
Fig.~\ref{fig:LC_longterm}, and created time-averaged GSC 
spectra in these individual bins.

We also investigated whether or not the source was detected 
with \maxi/GSC before the 2018 outburst. We created 72-day 
bin light curves in 3--4 keV, 4--10 keV, and 10--20 keV, 
applying the image fitting technique to the GSC data 
from 2009 September to 2016 July, in the same manner as those 
adopted in the \maxi~galactic and extragalactic X-ray 
source catalogs \citep{hor18, kaw18b}. 
We found, however, that the source was not detected 
significantly at any periods, 
and estimated 3$\sigma$ upper limits of the 7-year 
averaged fluxes as $\sim$0.2 mCrab in 4--10 keV, 
$\sim$0.1 mCrab in 3--4 keV, and $\sim$1.0 mCrab in 10--20 keV.

\subsubsection{\swift~data}
To acquire information at higher energies, we created  
hard X-ray spectra of MAXI J1820$+$070 from \swift/BAT-{\it survey} 
data. We processed the BAT survey data downloaded from 
the HEADAS 
archive\footnote{\url{https://heasarc.gsfc.nasa.gov/FTP/swift/data/obs/}} 
via the ftool {\tt batsurvey}, and then generated the 
time-averaged spectra and their response files 
in the individual continuous scans, using the 
script {\tt make\_survey\_pha}.
We selected the scans that partially or totally overlapped with 
the intervals of \maxi/GSC spectra, and used those scan data 
in the spectral analysis. 
If multiple BAT scans overlapping the interval of 
a \maxi~spectrum were present, we adopted the one 
with the longest 
overlapping time. If there are no overlapping scans, 
we discarded the \maxi~data. In the end, we obtained 
63 quasi-simultaneous \maxi/GSC and \swift/BAT-survey 
spectra, covering the 3--200 keV band. 

To investigate the more detailed spectral profile 
and X-ray variations on shorter time scales, 
we also analyzed the simultaneous \swift/XRT and 
BAT {\it event-by-event} data (hereafter we call 
BAT-event data), occasionally taken during 
the outburst. We picked out the observation IDs 
(OBSIDs) containing both XRT and BAT-event data 
(OBSID$=$00010627014, 00010627015, 00010627018
00010627026, 00010627035, 00010627036, 00010627045, 
00088657002, 00814259000, 00815603000), to create 
their light curves and spectra. 

The OBSID$=$00814259000 data were acquired 
on March 14, 1 day after the discovery with \maxi, 
and the other datasets were taken after March 19, 
when the source flux almost reached its peak level.
We found that all the datasets obtained after March 19 
had similar spectral and temporal properties and 
gave similar results. 
In the following, we just show the results from 
the data with OBSID$=$00814259000 (hereafter Data-1) 
and 00010627014 (Data-2), as representative data 
in the beginning of the outburst and at about 
the first flux peak, respectively. 
The former observation was performed from March 14 UT 
19:14:16 to 20:55:54, with net exposures of $\sim$1.0 ks 
for the XRT and $\sim$0.5 ks for the BAT, 
and the latter was from March 25 UT 04:07:30 
to 04:23:56, with a $\sim$2.4 ks exposure 
for the XRT and $\sim$0.5 ks for the BAT. 
The XRT was operated in the Windowed Timing mode 
in both observations. 

The XRT data were first reprocessed through the ftool 
{\tt xrtpipeline} with the calibration database (CALDB) 
downloaded in 2018 February. Then, the source signals 
were extracted from a circular region with a radius 
of 30 pixels, centered at the source position. 
To avoid the pileup effects, we excluded the 
events in the PSF core with a radius of 5 and 15 pixels 
for the Data-1 and Data-2, respectively,  
so that the count rate is well below 150 counts s$^{-1}$ 
\citep[see][]{eva09}. 
The background region of each dataset was 
defined as an annulus with inner and outer 
radii of 80 and 120 pixels, 
respectively, centered at the target position.
We employed {\tt swxwt0to2s6psf3\_20131212v001.rmf} 
in the CALDB as the XRT response matrix file. 
The ancillary response files (ARFs) of the 
individual observations were created via 
{\tt xrtmkarf} by considering the PSF profile. 

By using the Swift/BAT-event 
data, energy spectra and light curves were produced 
in the standard manner as described in \citet{sak07}. 
We created the energy response files with the ftool {\tt batdrmgen} 
and added systematic error vectors to the spectral files 
with {\tt batphasyserr}, before the spectral analysis.

\begin{figure}[ht!]
\plotone{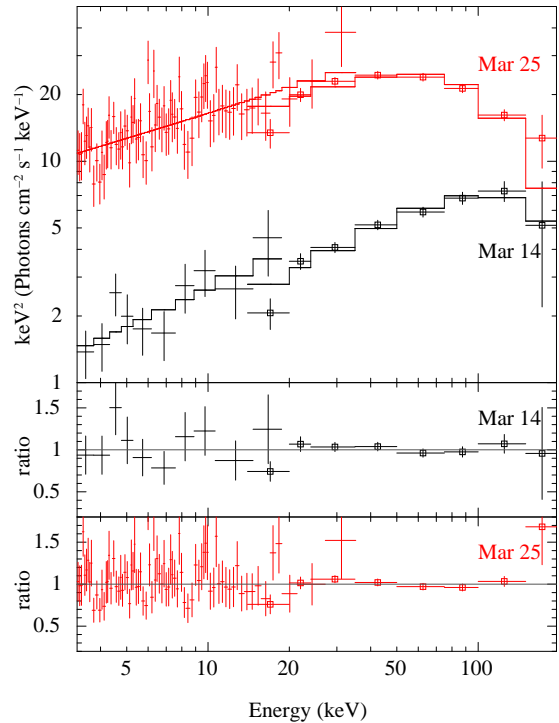}
\caption{\maxi/GSC (cross) and \swift/BAT (open square) 
spectra obtained on March 14 (black) and 25 (red), 
with their best-fit {\tt TBabs*nthcomp} models (top), 
and the data versus model ratios for the former (middle) 
and latter (bottom) spectra. The spectra are {\it unfolded} 
ones, corrected for the effective area of the instrument.
\label{fig:MAXIBATspec}}
\end{figure}

\subsection{Analysis and results}

\subsubsection{Long-term Evolution studied with \maxi/GSC and \swift/BAT-survey data} \label{sec:maxi_ana}
We first analyzed the \maxi/GSC and 
\swift/BAT-survey spectra to study the 
long-term spectral evolution, before the 
re-brightening in June. 
Figure~\ref{fig:MAXIBATspec} displays two 
typical spectra at low and high luminosities, 
obtained in 2018 March 14 
UT 9:42--17:44 
and March 25 UT 03:45-19:51, respectively. 
Both of them have a power-law-like  
profile, as usually seen in the low/hard 
state of BHBs. A significant spectral turnover 
can be seen in the March 25 spectrum at 
50--100 keV, which is not very clear in the 
March 14 spectrum.

\begin{figure}[ht!]
\plotone{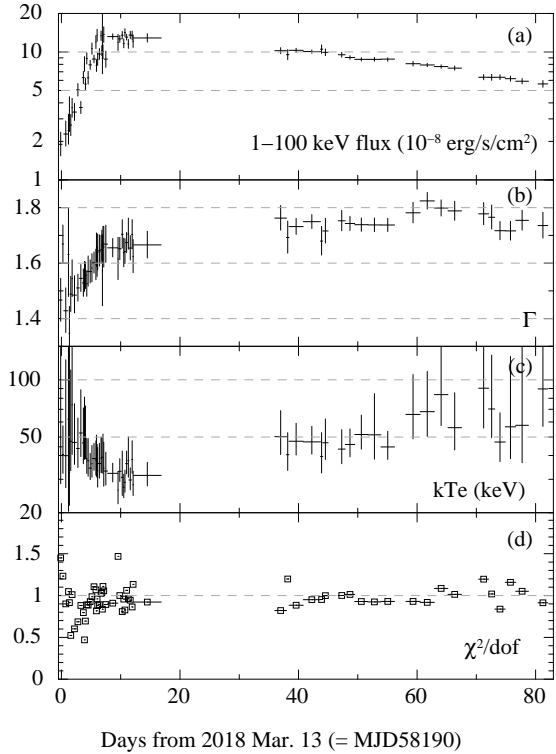}
\caption{Time variations of the parameters in the best-fit
{\tt TBabs*nthcomp} model. The unabsorbed 1--100 keV flux 
in units of $10^{-8}$ erg s$^{-1}$ cm$^{-2}$, 
the photon index, the electron temperature in units of keV, 
and the reduced chi-squared, from top to bottom. \label{fig:trend_fitpars}}
\end{figure}

We applied the Comptonization model {\tt nthcomp} 
to the individual GSC$+$BAT-survey spectra, 
assuming that the seed photons originate in 
the emission from the standard accretion disk. 
The {\tt nthcomp} model calculates a Comptonized  
spectrum from a photon index, an electron 
temperature, and a characteristic temperature 
of the seed photons (which is the inner disk 
temperature $\Tin$, in the case of the disk blackbody 
radiation). We fixed $\Tin$ at 0.1 keV, because 
it cannot be constrained in the GSC$+$BAT spectra, 
covering energies only above 3 keV. We have confirmed 
that the results was kept unchanged within the 90\% 
error ranges, when $\Tin=0.5$ keV and $0.05$ keV were 
adopted. To account for the interstellar absorption, 
we combined the {\tt TBabs} model \citep{wil00}, 
with a hydrogen column density of $1.5 \times 10^{21}$ 
cm$^{-2}$, which was determined from the {\it NICER} 
spectrum in March 12--14 \citep{utt18}. We varied the 
cross-normalization factor of the BAT data with 
respect to the GSC data. We obtained $\sim 1.0$ 
with a 90\% confidence range of $\pm \sim$20\% as its typical value.

The model well reproduced the spectra. 
In Fig.~\ref{fig:MAXIBATspec}, we show the 
best-fit models and the data versus model ratios 
of the March 14 and 25 data. 
Figure~\ref{fig:trend_fitpars} presents the overall trend in 
the fit parameters and the reduced chi-squared values.
During the outburst rise, the electron temperature 
and the photon index showed a slight decrease and 
increase, respectively. After the flux peak, 
the electron temperature increased slightly, 
whereas the photon index nearly unchanged 
within the error range, during the gradual decay of 
the unabsorbed 1--100 keV flux by a factor of $\sim$2.

\subsubsection{Short-term variation studied with \swift/XRT and BAT-event data} \label{sec:xrtbatana}
\begin{figure}[ht!]
\plotone{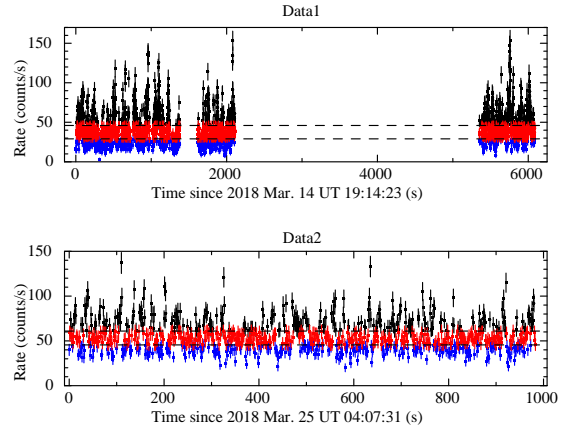}
\caption{\swift/XRT light curves of MAXI J1820+070 in 0.3--10 keV 
obtained from Data-1 (top) and Data-2 (bottom) with 1 s bins. 
The dashed lines indicate the thresholds to extract the 
intensity-sorted spectra in Figure~\ref{fig:fit_compps}, 
and the black and blue points are included in high 
and low flux phases, respectively. \label{fig:XRTlc}}
\end{figure}

We next analyzed the simultaneous \swift/XRT and 
BAT-event data occasionally acquired in the outburst.
Figure~\ref{fig:XRTlc} presents the XRT 1-s bin 
light curves of MAXI J1820$+$070 on March 14 and 25, 
obtained from Data-1 and Data-2, respectively. 
The flux varied by a factor of $\sim$2--5 
on timescales of a few to $\sim$100 s in both epochs. 
To investigate the energy dependence of the 
rapid flux variation, we sorted the time bins 
in these light curves, in terms of their count 
rates. In each observation, we defined the upper 
and lower 30\% time bins among all the data points 
as the high and low flux phases, respectively,  
and produced time-averaged spectra in 
these two phases (see Fig~\ref{fig:XRTlc} 
for thresholds of the count rates for these phases).

\begin{figure}[ht!]
\plotone{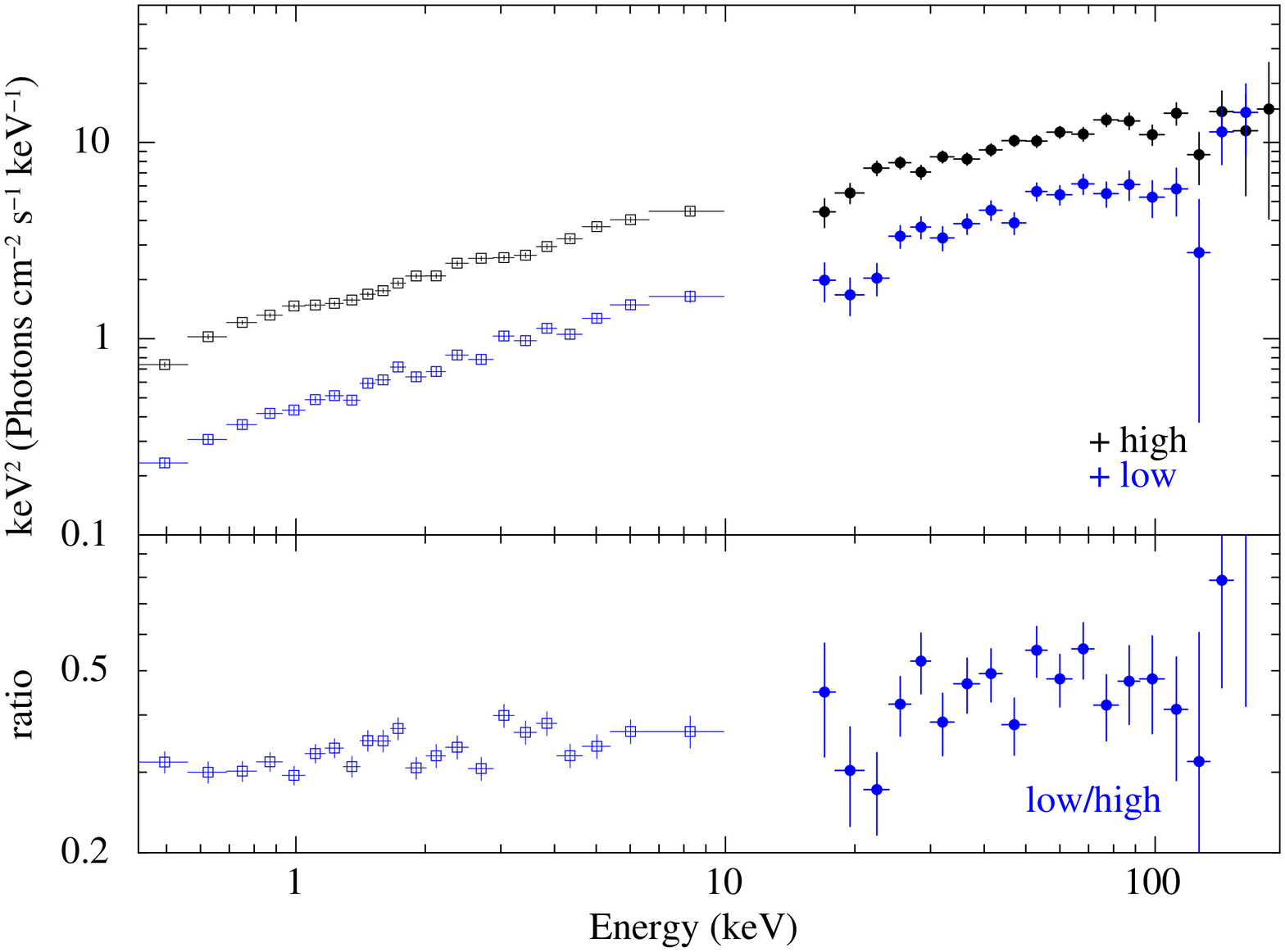}
\plotone{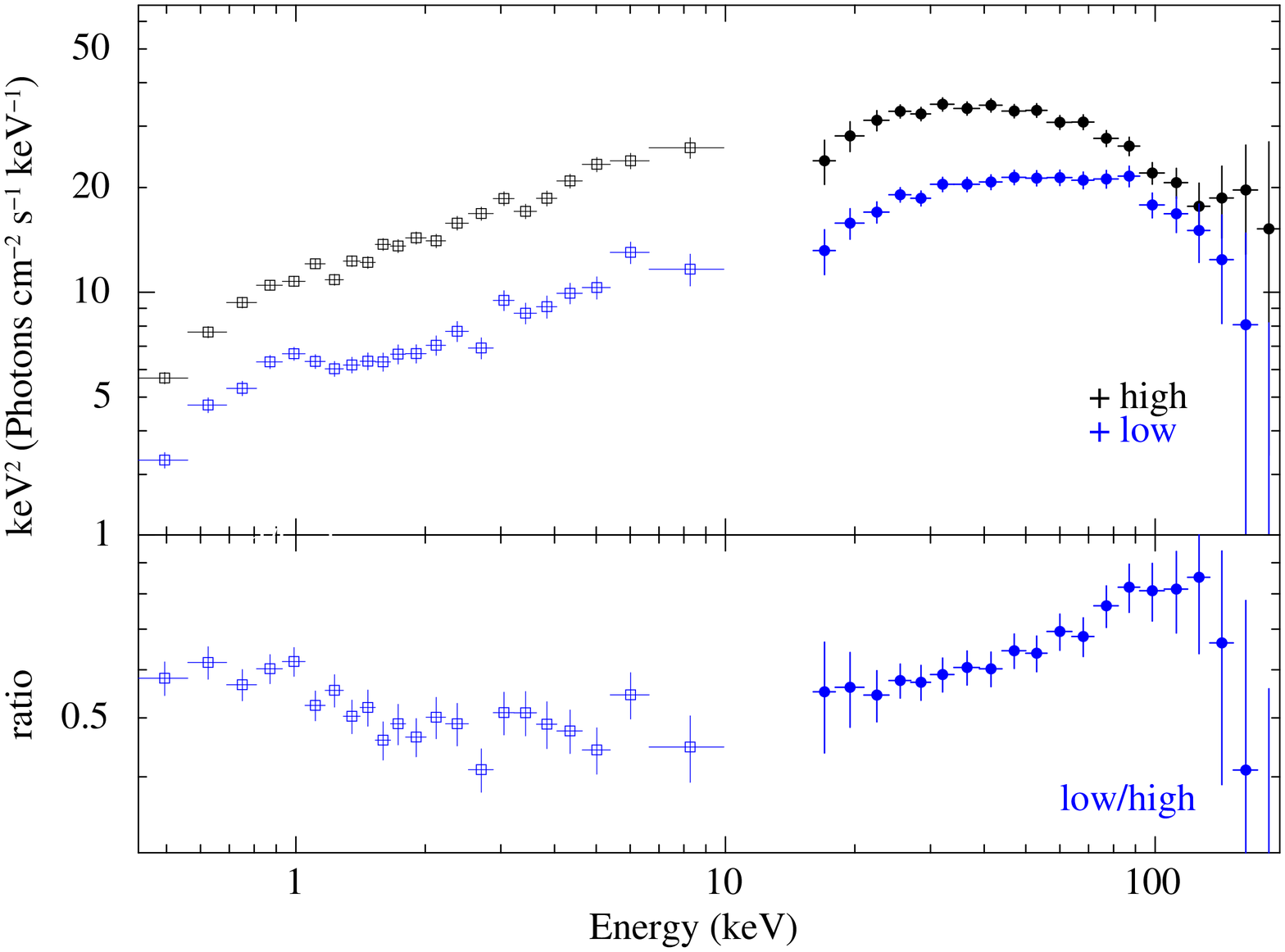}
\caption{(Top) Simultaneous \swift/XRT (open square) and 
\swift/BAT (filled square) intensity-sorted spectra 
on March 14 (Data-1; left) and 25 (Data-2; right). 
They are unfolded with a power-law model having 
a photon index of 2. (Bottom) Ratios 
of the unfolded spectra in the low intensity phases with 
respect to those in the high intensity phases. 
\label{fig:specratio}}
\end{figure}

\begin{figure*}[ht!]
\plottwo{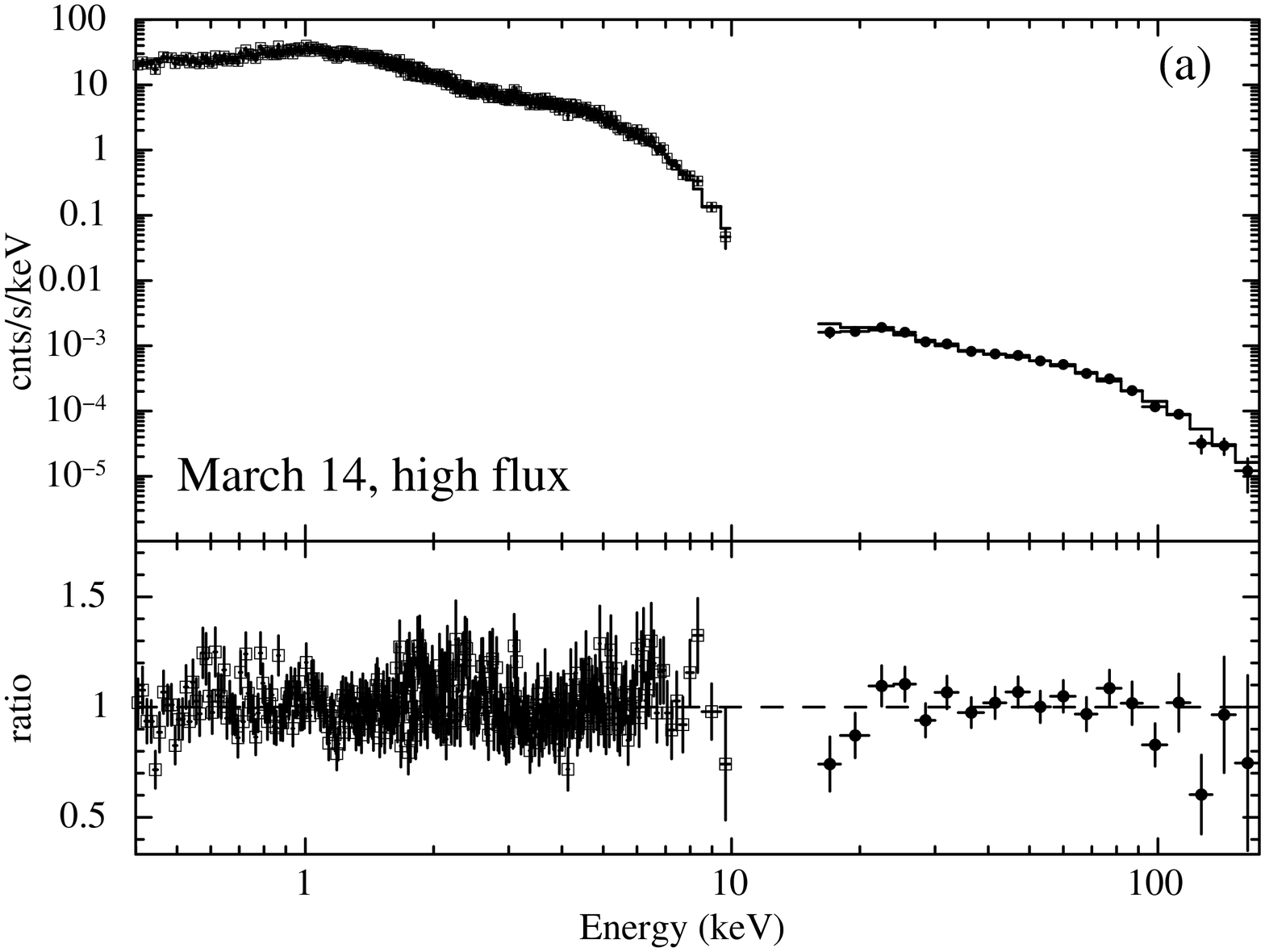}{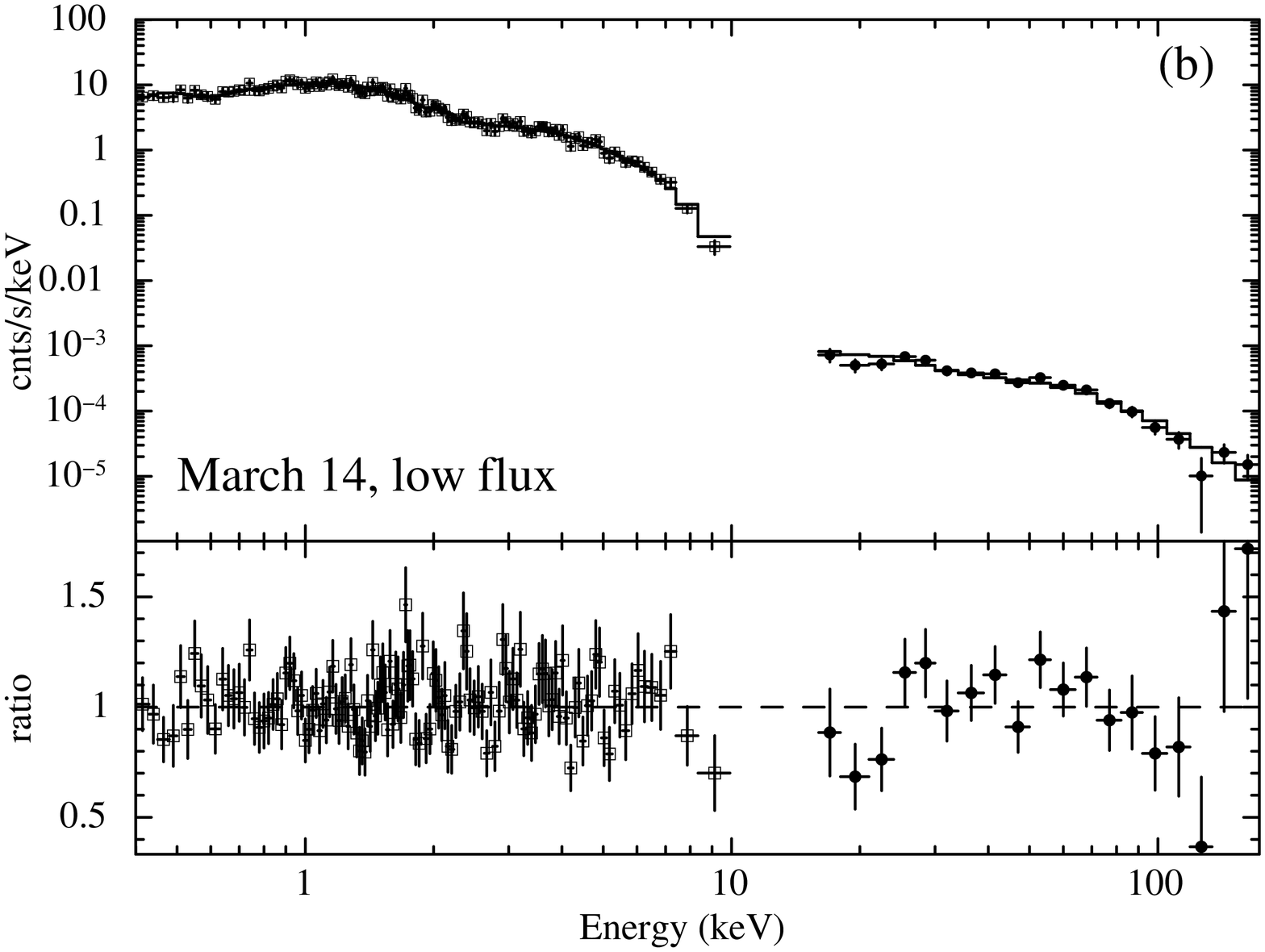}
\plottwo{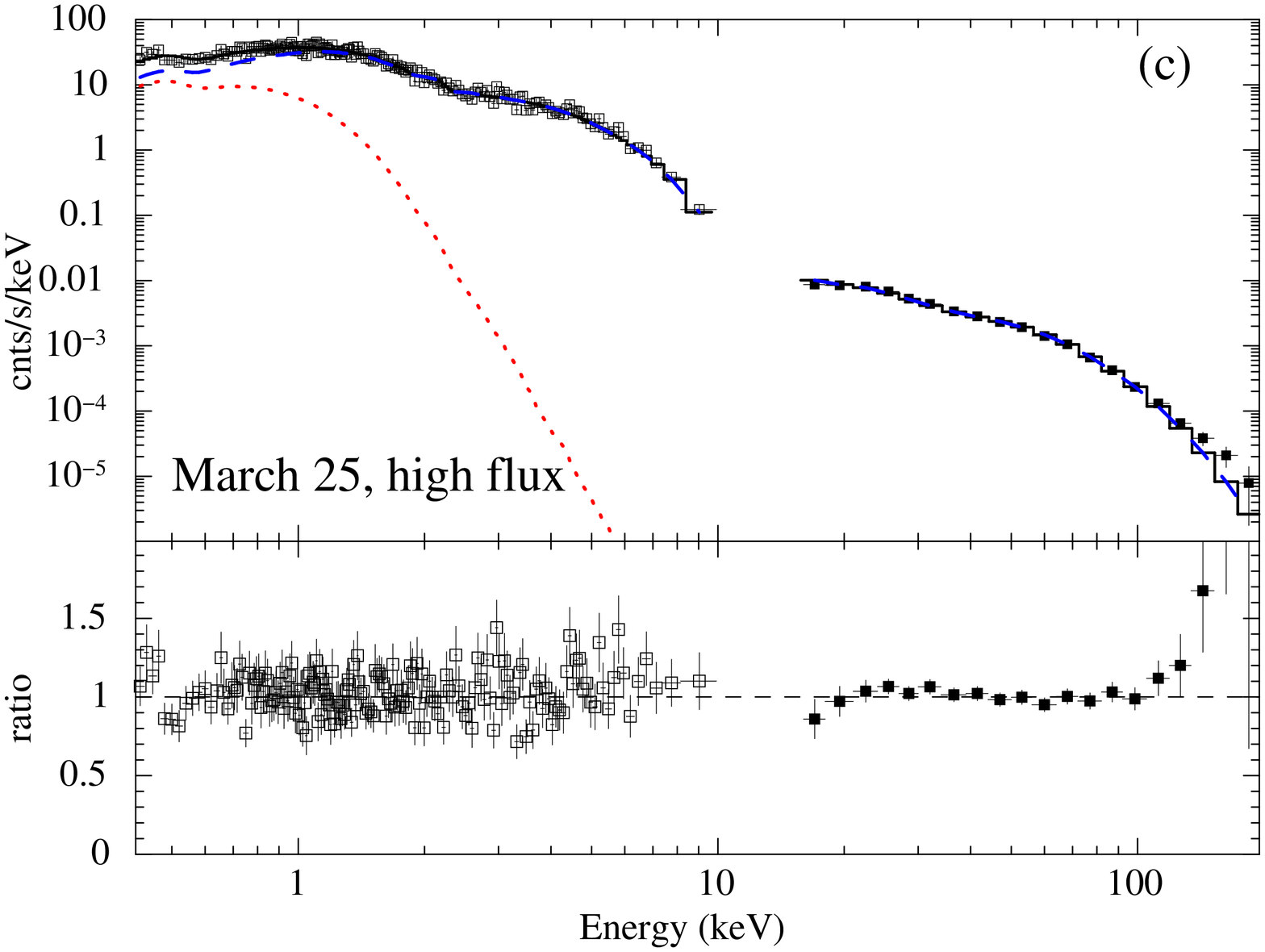}{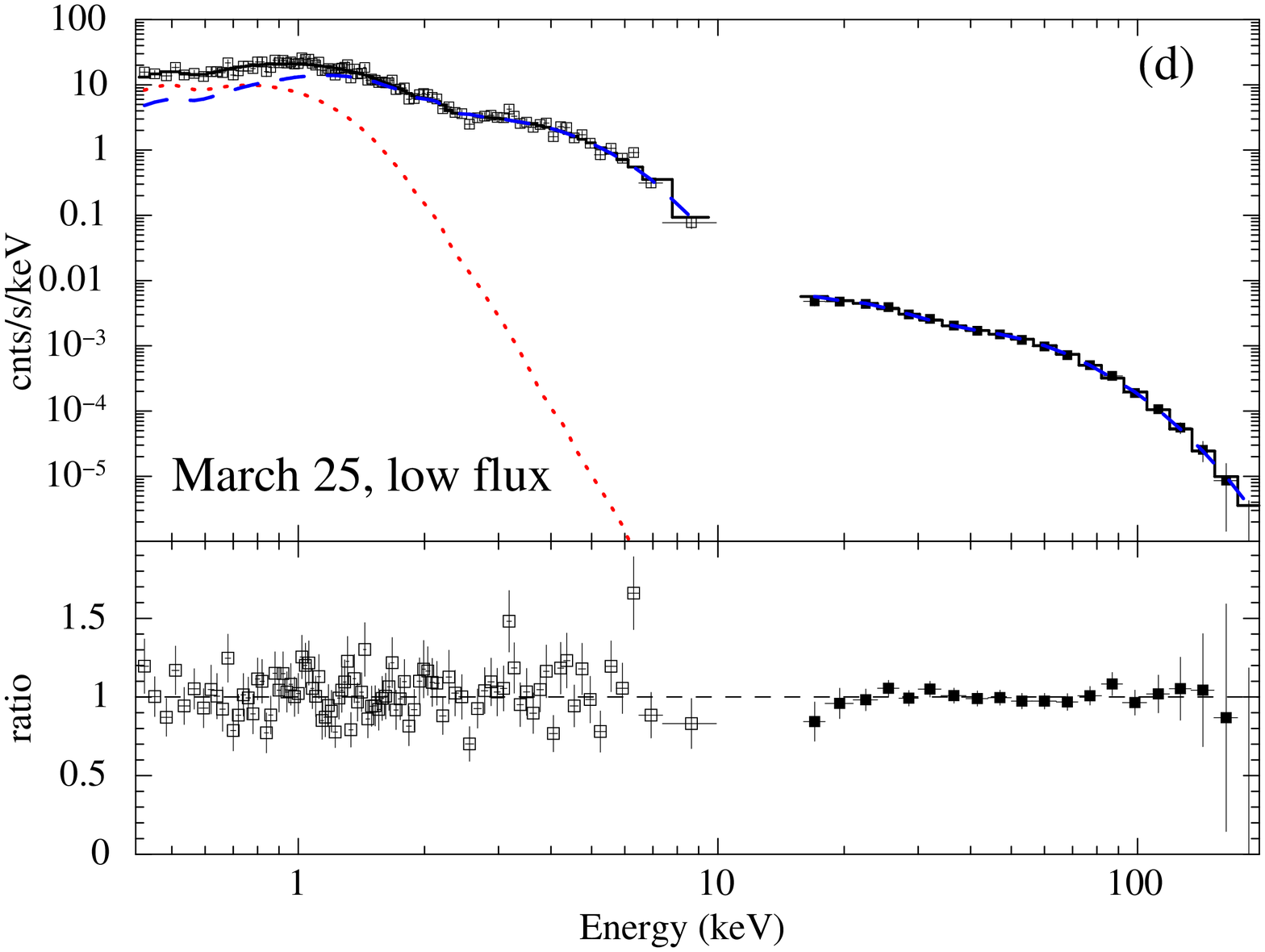}
\caption{(a),(b) the time-averaged folded spectra in the high and low 
intensity phases of Data-1, and their best-fit 
{\tt TBabs*compps} models. The lower panels present 
the data versus model ratio.
(c),(d) same as top panels, but for Data-2
with the best-fit {\tt TBabs*(disk+compps)} models. 
The {\tt diskbb} and {\tt compps} components are plotted 
separately with red dotted and blue dashed lines, respectively. 
\label{fig:fit_compps}}
\end{figure*}

\begin{deluxetable*}{clCCCC}[ht!]
\tablecaption{Best-fit {\tt TBabs*compps} and {\tt TBabs*(diskbb+compps)} parameters for Data-1 and Data-2, respectively, 
in the high and low intensity phases \label{tab:fit_compps}}
\tablecolumns{6}
\tablenum{2}
\tablewidth{0pt}
\tablehead{
\colhead{} &
\colhead{} &
\multicolumn{2}{c}{Data-1} 
& \multicolumn{2}{c}{Data-2} \\
\multicolumn{2}{c}{parameters} & \colhead{high} &
\colhead{low} & \colhead{high} & \colhead{low}
}
\startdata
{\tt TBabs} & \nh~($10^{22}$ cm$^{-2}$) & 0.10^{+0.05}_{-0.04} 
& <0.1 & 0.12 \pm 0.10 & 0.15^{+0.09}_{-0.07} \\
{\tt compps}\tablenotemark{a} & $kT_\mathrm{e}$ (keV) & >240 & >48 & 27^{+3}_{-2} & 35^{+8}_{-5} \\
& y-parameter & 1.1 \pm 0.2 & 1.6^{+0.1}_{-0.2} & 0.77 \pm 0.02 & 0.87^{+0.03}_{-0.02} \\
& $kT_\mathrm{bb}$ (keV) & 0.22 \pm 0.02 & 0.2^{+0.2}_{-0.1} 
& 0.18^{+0.12}_{-0.04} & 0.21^{+0.05}_{-0.04}\\
& norm ($10^4$)& $8^{+4}_{-3}$ & $1.3^{+6.8}_{-1.0}$ 
& $53^{+90}_{-27}$ & $17^{+22}_{-10}$ \\
{\tt diskbb}  & norm ($10^5$) & -  & - & 5.3^{+41.0}_{-5.1} & 4^{+14}_{-3} \\
& $r_\mathrm{in}$ (km)\tablenotemark{b} &  - &  - & 372^{+168}_{-198} & 240^{+192}_{-48} \\
\multicolumn{2}{l}{$\chi^2/$d.o.f.} & 356/328 & 168/144 & 200/190 & 108/107 \\
\multicolumn{2}{l}{flux\tablenotemark{c} ($10^{-8}$ erg s$^{-1}$ cm$^{-2}$)} & 4.1 
& 2.1 & 18 & 10 \\
\enddata
\tablenotetext{a}{The seed spectrum was assumed to be a disk blackbody, and the inner disk temperature of {\tt compps} ($kT_\mathrm{bb}$) was linked to that of {\tt diskbb} ($kT_\mathrm{in})$) in Data-2. The reflection component was ignored.}
\tablenotetext{b}{Inner radius estimated from the total photons of the disk blackbody emission, including the Comptonized photons in a spherical corona (see Section~\ref{sec:xrtbatana}). A distance and 
an inclination angle of 3 kpc and 30$^\circ$ are assumed, respectively. The color-temperature correction and the correction of the inner boundary condition are not considered.}
\tablenotetext{c}{Unabsorbed 0.01--100 keV flux.}
\end{deluxetable*}

In Figure~\ref{fig:specratio}, we plotted the 
time-averaged spectra in the low- and high-intensity 
phases, and the ratio of the low-intensity-phase 
spectra with respect to the high-intensity-phase 
spectra, produced from the individual datasets.
Both spectra in Data-1 can be approximated by a 
single power-law, and the spectral ratio increases with energy, 
indicating that the photon index in the low intensity
phase is slightly lower than that in the high-intensity 
phase. For Data-2, both spectra show a 
clear spectral cutoff at around 30--50 keV. 
Remarkably, the low-intensity-phase spectrum 
displays a hump at $\sim$1 keV, and the spectral 
ratio below $\sim$2 keV decreases with increasing 
energy, suggesting that a less variable component 
than the main cut-off power-law component 
is present in the soft X-ray band.

We applied a 
Comptonization model to these intensity-sorted 
spectra and investigated which physical 
parameter(s) made the spectral differences. 
Here we adopted a sophisticated Comptonization 
model, {\tt compps}, \citep{pou96} instead of {\tt nthcomp}. 
This model calculates a Comptonized spectrum produced 
in a hot electron cloud, based on exact numerical solutions 
of the radiative transfer equation, for a given  
electron temperature $kT_\mathrm{e}$, 
Compton y-parameter, geometry of the cloud, 
and the energy distribution of the seed photons. 
We assumed spherical geometry ({\tt geom} $=4$ 
in the XSPEC terminology) 
of the Comptonization component, and 
a multi-color disk blackbody as the seed spectrum. 
We ignored the reflection component, whose strength 
were not constrained, likely due to the uncertainties in 
the cross normalization between the XRT and BAT-event 
spectra. 

For Data-2, we combined the {\tt diskbb} model \citep{mit84} 
to {\tt compps}, as a direct disk blackbody 
component, to model the hump seen in the soft X-ray band.
The inner disk temperature of {\tt diskbb}, $kT_\mathrm{in}$, 
was linked to the seed temperature $kT_\mathrm{bb}$ 
of the {\tt compps} model. We also employed the 
{\tt TBabs} model as interstellar absorption, 
leaving $N_\mathrm{H}$ as a free parameter.
In this analysis, we varied the cross-normalization 
factor of the BAT with respect to the XRT, which 
was found to be consistent with 1.0, with a 90\% error 
of $\pm$ 0.1--0.2, in both Data-1 and Data-2.

These models, {\tt TBabs*(compps)} and 
{\tt TBabs*(diskbb\\+compps)}, successfully 
reproduced the Data-1 and Data-2 spectra, respectively. 
Figure~\ref{fig:fit_compps} shows the 
folded spectra with their best-fit models, 
and the data versus model ratio. Table~\ref{tab:fit_compps} 
lists the best-fit parameters of each phase. In both 
datasets, the Compton y-parameters in the low intensity
phases were larger than those in the high intensity phases.

We estimated the inner disk radius for Data-2,
from the photon fluxes of the direct disk 
component and the Comptonized component, via 
the equation given in \citep{kub04}: 
\begin{eqnarray*}
P_\mathrm{d} + P_{c} \cdot 2 \cos i = 0.0165 \left[ \frac{r_\mathrm{in}^2 \cos i}{(D/ 10~\rm{kpc})^2} \right] \left( \frac{kT_\mathrm{in}}{1~\rm{keV}} \right)^3  
\\~ \rm{Photons~s^{-1}~{cm}^{-2}},
\end{eqnarray*}
(where $P_\mathrm{c}$ and $P_\mathrm{d}$ are photon fluxes of the Comptonized component and the 
direct disk component, respectively), by assuming a spherical geometry of the 
Comptonization component and the conservation of the number of the disk 
photons after Comptonization.

\section{Optical and near-IR data}
\subsection{Observations and data reduction}
Optical and near-IR observations in the $g'$, $R_\mathrm{c}$, $I_\mathrm{c}$, $r$, $i$, $z$, $J$, $H$, and $K_\mathrm{s}$ bands were carried out with ground-based telescopes through the Target-of-Opportunity (ToO) 
program in the Optical and Infrared Synergetic Telescopes for Education and Research (OISTER).
The $g'$-, $R_\mathrm{c}$- and $I_\mathrm{c}$-band data were taken with the three-color imaging system developed for the MITSuME project \citep{kotani2005,yatsu2007,shimokawabe2008,2010AIPC.1279..466Y} on the MITSuME 50\,cm telescope in Akeno, the 50\,cm telescope at the Okayama Astrophysical Observatory (OAO) and the MURIKABUSHI 105\,cm telescope at the Ishigakijima Astronomical Observatory.
The $r$-, $i$- and $z$-band data were taken with Multi-wavelength SimultaneouS High throughput Imager and polarimeter (MuSaSHI), installed on the 55\,cm SaCRA telescope at Saitama University. The $J$-, $H$- and $K_\mathrm{s}$-band data were taken with the Nishi-harima Infrared Camera (NIC; \citealp{ishi11, tak13}) on the 2.0\,m Nayuta telescope at the Nishi-Harima Astronomical Observatory.
The data were reduced on IRAF by following standard procedures including bias and dark subtraction, flat fielding, and bad pixel masking. Photometry was performed with IRAF.
The magnitudes of MAXI J1820+070 were calibrated with nearby reference stars. 
The magnitudes of the reference stars were taken from the UCAC4 catalog \citep{2013AJ....145...44Z} for the $g'$-, $R_\mathrm{c}$- and $I_\mathrm{c}$-band data, from Pan-STARRS1 Surveys \citep{2016arXiv161205560C} for the $r$-, $i$- and $z$-band data, and from the Two Micron All Sky Survey Point Source Catalog \citep{2003tmc..book.....C} for the $J$-, $H$- and $K_\mathrm{s}$-band data. 
The statistical photometric errors of MAXI J1820+070 and the systematic errors of the reference star magnitudes were accounted for the observational errors. 

\begin{figure}[ht!]
\plotone{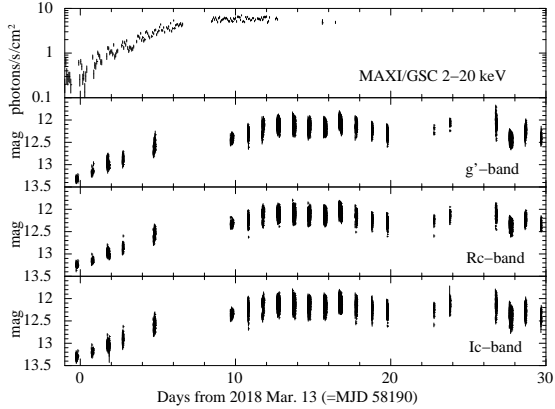}
\caption{X-ray and optical light curves of MAXI J1820$+$070, 
obtained with MAXI/GSC and in the MITSuME project, respectively.
The light curves in 2--20 keV, and in the optical $g'$, $R_\mathrm{c}$, 
and $I_\mathrm{c}$-bands are presented, from top to bottom. 
The magnitudes are expressed in the AB system. 
The error bars of the optical light curves include both the statistical 
photometric errors of MAXI J1820+070 and the systematic errors of the 
reference star magnitudes. 
\label{fig:optlc}}
\end{figure}

Figure~\ref{fig:optlc} shows the $g'$, $R_\mathrm{c}$, $I_\mathrm{c}$-band 
light curves of MAXI J1820$+$070, together with 
the MAXI/GSC light curve in 2--20 keV. 
The optical fluxes were found to gradually increase  
as the X-ray flux became higher, and show significant 
variation in each night by 0.5--1 mag 
(by a factor of 1.6--2.5, in the flux unit).
As described in the following section, we studied the 
averaged properties of the multi-wavelength spectral 
energy distribution (SED) around the flux peak, 
by combining the one-night averaged optical and 
near-IR data, and the quasi-simultaneous X-ray 
data obtained with \maxi/GSC and \swift/BAT. 
To investigate the short-term variations of 
the optical fluxes, and their correlation to 
those in X-rays, are left as a future work.

We also searched the archival optical and 
IR data\footnote{at \url{http://irsa.ipac.caltech.edu/frontpage/}.}
for possible activities of \srcname before the 2018 outburst. 
The source was detected multiple times by Palomar Transient Factory 
(PTF) in 2013 May with $R$-band magnitudes 
of 18.3--18.7 mag and typical errors of $\sim$0.06 mag. 
It was also detected in the mid-IR band 
with Wide-field Infrared Survey Explorer 
(WISE)/Near-Earth Object WISE (NEOWISE)
in 2010 September/2014--2017 March and September. 
The WISE/NEOWISE apparent magnitudes were 15.3--16.3/13.7--14.8 mag 
with typical errors of $\sim$0.3/0.08 mag in the 
W1 band (3.4 $\mu$m), and 
14.7--15.4/13.4--14.6 mag with typical errors of 
$\sim$0.3/0.2 in the W2 
band (4.6 $\mu$m). Both the W1 and W2 band fluxes showed 
significant variations by 0.5--1 mag 
(a factor of $\sim$2 in the flux unit) 
within a few to several days.

\subsection{Analysis of Multi-wavelength SEDs} \label{sec:SED}

\begin{figure*}[ht!]
\plottwo{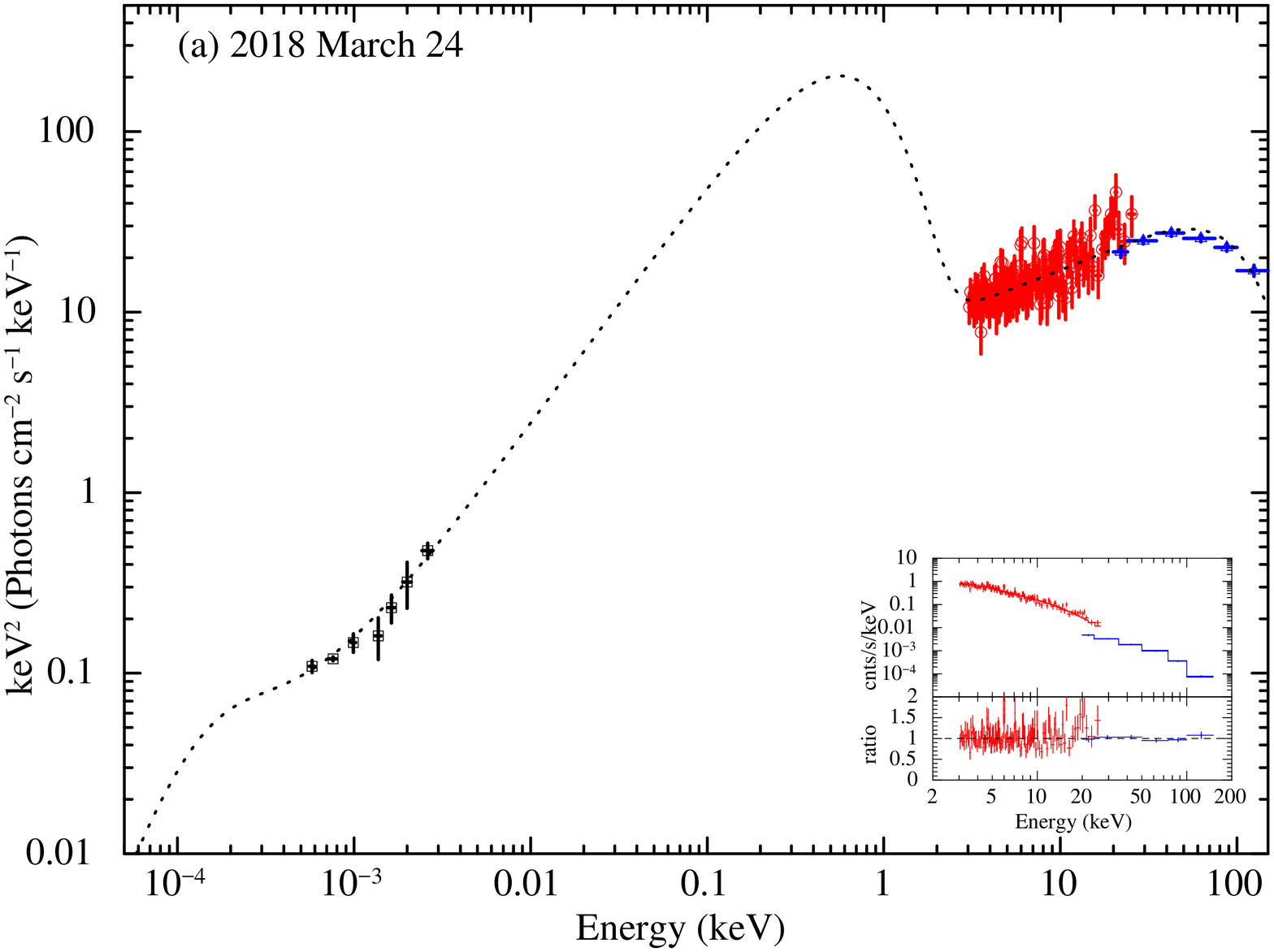}{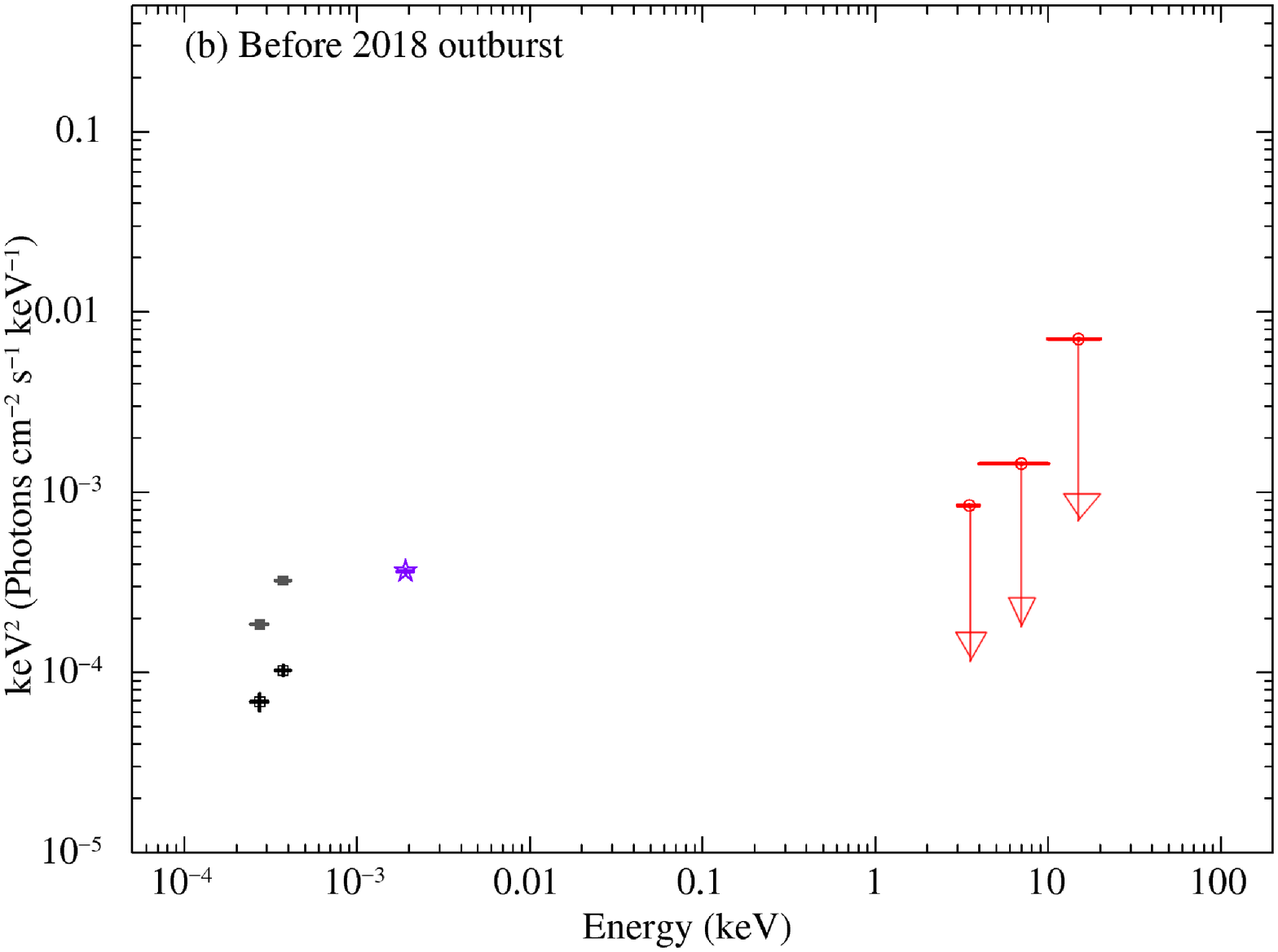}
\caption{(a) Multi-wavelength SED of MAXI J1820$+$070 on 2018 March 24. 
The black squares indicate the optcal/IR data in the $K_\mathrm{S}$, 
$H$, $J$, $z$, $i$, $r$, and $g'$ bands, from 
left to right. Red circles and blue triangles show the MAXI/GSC 
and Swift/BAT data, respectively. The best-fit {\tt diskir} model 
is presented in the dotted line. The inset presents the same X-ray 
data folded with the instrumental responses, in units of counts s$^{-1}$ keV $^{-1}$ (upper), and their data 
versus model ratio (lower). (b) SED before the start of the 
2018 outburst, obtained from the archival PTF $R$-band 
data (purple star), and the WISE (black open squares)/NEOWISE 
(grey filled squares) IR data, combined with the 
3$\sigma$ upper limits of the 7-year averaged X-ray fluxes 
derived with the MAXI/GSC (see Section~\ref{sec:gsc_reduction}). 
\label{fig:SED}}
\end{figure*}

Figure~\ref{fig:SED}(a) shows the multi-wavelength SED 
on 2018 March 24, when the $g'$, $r$, $i$, $z$, $J$, $H$, $K_\mathrm{s}$-band 
data, together with the X-ray data from \maxi/GSC and 
\swift/BAT, were available. 
The X-ray data were corrected for interstellar absorption 
using the {\tt TBabs} model with $N_\mathrm{H} = 
1.5 \times 10^{21}$ cm$^{-2}$ \citep{utt18}, and optical/IR data 
for interstellar extinction using the {\tt redden} model  
in XSPEC with the extinction $E(B-V) = 0.26$, which was 
converted from the $N_\mathrm{H}$ value through the 
relation in \citet{boh78}. 
As noticed from the figure, the SED in the IR 
to optical band is not smoothly connected to the 
X-ray spectrum. 
For comparison, we also plotted, in 
Figure~\ref{fig:SED}(b), an SED obtained 
from the archival optical/IR data before 
the 2018 outburst, and the upper limits 
of the X-ray fluxes estimated with the MAXI/GSC.

\begin{figure}[ht!]
\plotone{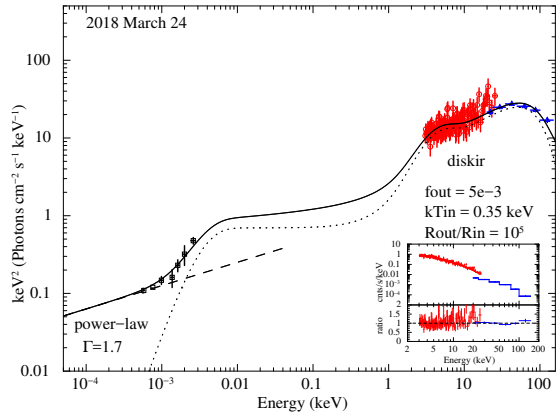}
\caption{Same data as Fig.~\ref{fig:SED}(a), with the  
{\tt powerlaw+diskir} model (solid line). 
The power-law and {\tt diskir} components are presented in the 
dashed and dotted lines, respectively.\label{fig:SEDjet}}
\end{figure}

The optical and IR photons of BHBs can originate in the 
accretion disk, jet, or companion star. In the 
case of the disk emission, the optical and IR bands 
corresponds to the radiation from the outer disk 
region, which is often irradiated by the X-rays 
from the inner disk region. 
To obtain the upper limit of the contribution 
of the disk emission to the optical/IR fluxes, 
we applied the irradiated disk model 
{\tt diskir} \citep{gie08,gie09} to the multi-wavelength SED on 
2018 March 24 (see section~\ref{sec:discussion_optir} 
for discussion of the contributions of 
the other emission components).

The {\tt diskir} model computes the spectrum of the 
disk emission and its Comptonization, considering the 
irradiation of the disk. The model has 9 input parameters: 
the inner disk 
temperature $kT_\mathrm{in}$, the photon index 
$\Gamma$ and electron temperature $kT_\mathrm{e}$ 
of the Comptonized component, the luminosity ratio $L_\mathrm{c}/L_\mathrm{d}$ 
of the Comptonized component to the disk component, 
the fraction $f_\mathrm{in}$ of the luminosity of 
the Comptonzed component that is thermalized  
in the inner disk, the fraction $f_\mathrm{out}$ 
of the bolometric flux illuminating the outer disk, 
the radius $r_\mathrm{irr}$ of the inner disk 
irradiated by the Comptonized component with 
respect to the inner disk radius, 
the outer disk radius $R_\mathrm{out}$,
and the normalization, determined by 
the inner disk radius $R_\mathrm{in}$ 
in the same manner as {\tt diskbb}. 

Following previous works \citep[e.g.,][]{shi17}, 
we fixed $r_\mathrm{irr}$ and $f_\mathrm{in}$, which 
were unconstrained with our data, 
at default values, 1.1 and 0.1, respectively, 
which are appropriate for the low/hard state 
spectra \citep{pou97}.
The other parameters were left as free parameters.
We multiplied {\tt TBabs} with $N_\mathrm{H}$ fixed at 
$1.5 \times 10^{21}$ cm$^{-2}$ and {\tt redden} with 
$E(B-V)$ at $0.26$ to the {\tt diskir} model, 
as X-ray interstellar absorption and optical/IR extinction, 
respectively. We have confirmed, however, that 
the conclusions below do not change if we increase 
and decrease $N_\mathrm{H}$ by a factor of 2, 
and change $E(B-V)$ accordingly.

As shown in Fig.~\ref{fig:SED}(a), the model was 
able to reproduce the SED on March 24, giving a 
$\chi^2$/d.o.f. value of $106/122$. In this best-fit 
model, the reprocessed emission from the outer 
disk dominates the flux in the IR bands, 
where a profile flatter than the optical 
SED is seen. 
The parameter values were constrained to be 
$kT_\mathrm{in} = 0.23^{+0.03}_{-0.13}$ keV, 
$\Gamma = 1.67^{+0.01}_{-0.03}$, 
$kT_\mathrm{e} = 29^{+2}_{-3}$ keV, 
$L_\mathrm{c}/L_\mathrm{d} = 0.24 \pm 0.06$, 
$f_\mathrm{out} = 6.4^{+4}_{-2} \times 10^{-5}$, and 
$R_\mathrm{out} > 3 \times 10^4 R_\mathrm{in} 
= 2 \times 10^7$ ($D$/3~kpc) 
($\cos i/\cos 30^\circ$)$^{-1/2}$ km. The 
$\Gamma$ and $kT_\mathrm{e}$ values were consistent 
with those obtained in the analysis of the X-ray 
spectra alone, and the $kT_\mathrm{in}$ value was 
the same as that determined from the \swift/XRT and 
BAT-event data on March 25, within the 90\% error 
ranges.

\section{Discussion}
\subsection{Overall X-ray Properties} \label{sec:discussion_xraylc}
Like other X-ray novae, the new X-ray source 
MAXI J1820$+$070 displayed a rapid flux rise, 
up to as high as $\sim$2 Crab in 2--20 keV, 
and then a slower 
decay for $\sim$3 months. 
Using MAXI/GSC data, we found that the source 
increased its X-ray flux $\gtrsim$4 orders of 
magnitude from the quiescent level, to 
$\sim$2 Crab in 2--20 keV band, at the peak 
in the end of 2018 March. 
Before the re-brightening in the middle of June, 
the source did not show any drastic spectral 
softening, and kept presenting a hard spectrum 
roughly characterized with an exponentially cut-off 
power-law model with a 
photon index of $\sim$1.5 and a cut-off energy of 
$\sim$50 keV, which is consistent with those 
of black hole X-ray binaries in the low/hard state.

The observed peak flux of \srcname was quite high, 
compared with typical flux levels at which many known 
BHBs in the Galactic center regions 
show the transition to the high/soft state 
\citep[$\sim$0.1 Crab; e.g.,][]{yu07,zho13}. 
This can be explained if \srcname is located 
closer to us than them. Indeed, $D \approx 3$ kpc has recently 
been obtained from the astrometry with Gaia data \citep{gan18b}. 
Using the unabsorbed 1--100 keV flux, 
obtained with the best-fit {\tt nthcomp} model 
of the simultaneous \maxi/GSC and 
\swift/BAT spectrum on 2018 March 25, 
the peak luminosity in the end of March is 
calculated to be $\sim 2 \times 10^{38}$ ($D$/3 kpc)$^2$ erg s$^{-1}$, 
which is converted to an Eddington ratio of 
$L_\mathrm{X}/L_\mathrm{Edd} \sim 0.1$ ($D$/3 kpc)$^2$ 
$(M_\mathrm{BH}/10 M_\sun)^{-1}$. 

After the low/hard state period for $\sim 3$ months, 
the source caused re-brightening and entered into 
the state transition, at a luminosity similar to 
the first peak in March. Such a long period before 
the state transition is unusual in transient BHBs, 
although similar behavior was observed in 
the 2009 outburst of XTE J1752$-$223 \citep{nak10}, 
where the source stayed in the low/hard state 
for 3 months before the transition, with two plateau 
phases in its X-ray lightcurve. In that case, the complex evolution 
was explained by a gradual increase of the mass accretion 
rate for some unknown reasons. 

What made the double-step rise of 
\srcname is still unclear, 
but possibly the first rise was caused due to an 
enhancement of the mass transfer from the companion 
star, and the second rise due to a rapid increase 
of the mass accretion rate caused by the disk 
instability that was triggered somewhere in the 
outer disk during the first rise and propagated 
inward. The viscous timescale of the disk is 
expressed as 
$t_\mathrm{v}(R) \sim \alpha^{-1} \Omega_\mathrm{K}^{-1} (H/R)^{-2}$, 
where $\alpha$, $H$, and $\Omega_\mathrm{K}$ are the viscosity 
parameter, the disk scale height, and the Keplerian angular 
velocity at the radius $R$, respectively. The time scale 
$t_\mathrm{v} \sim 90$ days corresponds to $R \sim 2 \times 10^{10}$ cm 
for $H/R \sim 0.01$ and $\sim 4 \times 10^{11}$ cm for $H/R \sim 0.1$, 
when a black hole mass of 10 $M_\sun$ and $\alpha = 0.1$ are assumed.

An alternative interpretation for the two-step flux increase 
may be provided in terms of the irradiation 
of the companion star, as invoked by \citet{nak14} to explain the 
re-flare observed in the outburst of Swift J1910.2$-$0546 
(or MAXI J1910$-$057). The first flux rise could be produced by 
the enhancement of the mass accretion rate through 
the inner disk due to the disk instability, and 
that the strong X-rays irradiated and inflated 
the companion star, causing increase of the gas supply 
to the accretion disk. The second flux 
enhancement could then be realized by 
triggering the disk instability again.

\subsection{Implications in Long- and Short-term X-ray Variations}

Looking at the \maxi/GSC and \swift/BAT spectra during the 
low/hard state in more detail, we found that the spectrum 
became slightly softer 
and bent at lower energies at higher luminosities. 
According to the best-fit {\tt nthcomp} models for the 
individual spectra (Section~\ref{sec:maxi_ana}), the 
photon index increased from 1.4 to 1.7 and the electron 
temperature decreased from 50 keV to 30 keV, during the  
rise phase of the outburst. This long-term spectral 
evolution would be explained by the change in the mass 
accretion rate; the standard disk is developed 
inwards as the mass accretion rate increases, and the 
soft X-rays from the standard disk cools the hot 
inner flow and/or corona around the standard disk, 
providing a softer Comptonized spectrum, with a lower 
electron temperature.

We also detected spectral variation on much shorter 
timescales, a few to $\sim$100 s, using the 
\swift/XRT and BAT-event data taken on March 14 and 25, 
corresponding to the beginning and the peak of the 
outburst. In both periods, the spectrum of \srcname was 
softer/harder at high/low intensity 
phases of the short-term variation. Applying 
the {\tt compps} model, 
a smaller y-parameter (and a smaller electron 
temperature on March 25) 
were obtained at higher flux phases.
Similar trends were obtained 
in the shot analysis of Cyg X-1 \citep{neg94, neg01, yam13a}, 
and the density fluctuation in the radiation inefficient 
accretion flow was suggested as one possibility to 
drive the variation, on the basis of the evolution 
of the spectrum and X-ray time lags during the shots. 
Further studies of timing properties of \srcname would 
be required to understand the actual cause of its 
short-term variation.

Thanks to the good statistics of the \swift/XRT data, 
we detected, in the March 25 spectrum, 
a structure below $\sim$2 keV that cannot 
be reproduced by the {\tt TBabs*nthcomp} model 
and is less variable than the main Comptonization 
component. Assuming it as the direct disk emission component, 
we obtained the inner disk temperature as $\sim$0.2 keV and 
the inner disk radius as $\sim$180--540 ($D$/3 kpc) 
$(\cos i/\cos 30^\circ)^{-1/2}$ km, which was 
estimated from the photon flux of the total 
intrinsic disk emission component, including 
both the direct and Comptonized ones. 
This radius can be converted to 
$\sim$12--36 $R_\mathrm{g} (M_\mathrm{BH}/10$ $M_\sun)^{-1}$ 
($D$/3 kpc) $(\cos i/\cos 30^\circ)^{-1/2}$, where 
$R_\mathrm{g} = GM_\sun/ c^2$, 
and this means that the standard disk 
is truncated, around the flux peak in the end of March. 

We cannot rule out, however, the possibility that 
the soft component seen on March 25 was not the 
direct standard disk emission but a 
Comptonized emission produced around 
the inner edge of the standard disk, 
as discussed \citep{chi10, yam13b, shi14}. 
If this is the case, the direct disk component was  
below the energy range of the XRT, and the standard 
disk was further truncated, with a lower inner disk 
temperature than what we estimated above. 
The mass accretion rate at the flux peak can then be
close to the Eddington rate, considering the 
radiation efficiency of the standard disk, 
$\sim 0.1 R_\mathrm{ISCO}/R_\mathrm{in}$ (where 
$R_\mathrm{in}$ and $R_\mathrm{ISCO}$
the radii of the inner edge of the standard 
disk and the innermost stable circular orbit, 
respectively), and the peak luminosity 
estimated in Section~\ref{sec:discussion_xraylc}.

\subsection{Origin of Optical and Near-IR Emission in the Outburst and Jet Energetics}
\label{sec:discussion_optir}

The optical and near-IR fluxes of BHBs is 
considered to originate in the blackbody emission 
from the companion star, jet emission, and/or the 
emission from the outer region of the accretion 
disk, which is often enhanced by the irradiation 
of the X-rays from the inner disk region. In the 
case of \srcname during the outburst, 
the contribution of the companion star is negligible, 
because the previous PTF data suggest that the optical 
flux in the quiescent phase was at least $\sim$3 
orders of magnitude smaller than in the peak 
of the outburst (see Section~\ref{sec:SED}). 
The multi-wavelength SED around the X-ray flux peak 
was technically able to be fit with an 
irradiated disk model. However, the resultant 
value of $f_\mathrm{out}$, the strength of 
the reprocessed component, was 
unusually small compared with those of typical 
BHBs in the low/hard state \citep[$>10^{-3}$; e.g.,][]{gie09}. 

Considering the above results, we suggest that the optical 
and near-IR emission of \srcname in the outburst 
was not entirely produced by the disk emission, 
but substantially contributed by the jet emission, 
particularly in the near-IR band. Indeed, as shown in 
Figure~\ref{fig:SEDjet}, the SED can be fairly well 
reproduced, for example, by adding a power-law component (as the 
optically-thin synchrotron emission from jet) with 
a photon index of 1.7 and a normalization adjusted to the 
$K_\mathrm{S}$ band flux, and setting $f_\mathrm{out}$, 
$kT_\mathrm{in}$, $L_\mathrm{c}/L_\mathrm{d}$, and 
$R_\mathrm{out}$ of {\tt diskir} to be $5 \times 10^{-3}$, 
0.35 keV, 70, and $10^{5} R_\mathrm{in}$, respectively. 
In this model, the inner disk region 
is efficiently irradiated by the Comptonization component 
dominating the X-ray luminosity, and the heated disk produces 
a weak hump seen around $\sim 5$ keV. We note that the 
parameters related to the irradiation of the outer disk 
do not significantly change by the irradiation efficiency of 
the inner disk alone, because they are determined from the 
bolometric flux. Remarkably, the stronger contribution of 
the jet component at longer wavelengths, indicated from this 
model, is consistent with the observed sub-second 
optical variations \citep{gan18a} likely originating in 
the jet activity, which was found to be stronger in redder bands.

Indeed, previous studies of BHBs suggest that steady 
compact jets are present during the low/hard state, 
and that their emission dominates the radio to IR or 
optical fluxes \citep[see][and references therein]{fen10,mar10,gal10}. 
The SEDs of BHBs in the low/hard state exhibit a flat, 
power-law profile at the radio frequencies 
\citep[e.g.,][]{cor00, fen01}, with a 
spectral index of $\beta \sim0$ (where the flux density 
$F_\nu \propto \nu^\beta$), 
and have a break in the sub-mm to IR band, 
above which a smaller $\beta$ is obtained \citep{cor02,mig10,gan11,shi11,rus13,rus14}.
As discussed also for AGN jets \citep{bla79}, 
this SED profile can be described with the synchrotron 
radiation from conical jets, 
where optically-thick and thin synchrotron components are 
observed below and above the break, respectively. 

Such a broken-power-law shaped SED, extending from the radio 
to near-IR band, was actually obtained in \srcname \citep{rus18} 
around the X-ray flux peak at the end of March, with spectral 
indices of $\beta \sim 0.3$ and $\sim -0.7$, below and above 
the break frequency $\nu_\mathrm{b}$ of $\sim 3 \times 10^{13}$ 
Hz. Following \citet{shi11}, we attempt here to estimate 
the physical parameters of the jet base, from 
$\nu_\mathrm{b}$ and the flux density at 
$\nu_\mathrm{b}$ of $F_{\nu_\mathrm{b}} \sim 400$ mJy. 
For simplicity, we assume a single-zone jet base and 
ignore relativistic beaming effects by the bulk motion of the jet.
If the electron number density at the Lorentz factor $\gamma$ 
is proportional to $\gamma^p$, the synchrotron luminosity 
in an optically-thin part, $\nu L_\nu$ depends on 
$\nu$, $p$, the magnetic field strength $B$ and the 
volume $V$ of the emission region (where $V = 4\pi R^3/3$ 
in the case of a spherical region), and the pitch angle 
$\theta$ of the jet, while the synchrotron 
self-absorption coefficient $\alpha_\nu$ is expressed as 
a function of $\nu$, $p$, $B$, and $\theta$ 
\citep[see][for complete expressions]{shi11}.
The former parameter, $\nu L_\nu$, is proportional 
to $\nu^{(3-p)/2}$, and thus we obtain $p = 2.4$ 
for \srcname, from the observed spectral index, $\beta \sim -0.7$. 

Assuming the equipartition of the magnetic field 
energy and the kinetic energy of electrons in the jet, 
and considering the condition of the optical depth and 
the luminosity at $\nu_\mathrm{b}$ as $\sim \alpha_{\nu_\mathrm{b}} 
R \sim 1$ and $\nu_\mathrm{b} L_{\nu_\mathrm{b}}$ 
as $\sim 1 \times 10^{35}$ ($D$/3 kpc)$^2$ erg s$^{-1}$, respectively, 
we obtain $B = 1 \times 10^4$ ($D$/3 kpc)$^{-0.22}$ 
($\sin \theta/ \sin 30^\circ$)$^{0.55}$ G and 
$R = 2 \times 10^9$ ($D$/3 kpc)$^{0.94}$ 
($\sin \theta/\sin 30^\circ$)$^{0.11}$ cm 
for the jet base of \srcname. 
These values are comparable to those estimated in 
GX 339$-$4 \citep{shi11}, XTE J1550$-$564 \citep{cha11}, 
and MAXI J1836$-$194 \citep{rus14} during the low/hard state.

The magnetic energy density is derived from the above 
$B$ value as $u_B = B^2/8 \pi$ as $\sim 8 \times 10^6$ 
erg cm$^{-3}$. The Lorentz factor of electrons emitting 
$\nu_\mathrm{b} = 3 \times 10^{13}$ photons is $\sim 10$.
Following \citet{cha11}, we can calculate the 
timescales of adiabatic cooling and radiative cooling 
at the jet base as $\gtrsim B/c \sim 70$ ms 
and $\propto u_B^{-1} \gamma^{-1} \sim 400$ ms, 
respectively. This indicates that the former 
is the dominant cooling process.

The total synchrotron luminosity 
$L_\mathrm{sync}$ is roughly estimated as $\sim 10^{36}$  
erg s$^{-1}$, from the SED profile obtained in \citep{rus18} and 
the energy density of the synchrotron radiation, $u_\mathrm{sync} 
\sim L_\mathrm{sync} / (4 \pi R^2 c)$, 
is calculated to be $3 \times 10^{5}$ erg cm$^{-3}$.
The luminosity of the synchrotron self-Compton radiation, 
$L_\mathrm{SSC}$, is thus estimated as 
$L_\mathrm{sync} u_\mathrm{sync} / u_B \sim 0.05 L_\mathrm{sync}$. 
This suggests that the synchrotron self-Comptonization 
emission is negligible, contributing only $\sim$0.05\% 
to the X-ray flux. 
The energy density of external photons from the 
accretion disk is roughly estimated as $9 \times 10^7$ erg cm$^{-3}$, 
from the 1--100 keV X-ray luminosity, $\sim 1 \times 10^{38}$ 
erg s$^{-1}$. The contribution of the external Comptonization 
emission is thus $\sim 1 \times 10^{37}$ erg s$^{-1}$, which is 
still only $\sim$10\% of the total X-ray luminosity.
As noticed in figure~\ref{fig:SED}(a), the observed 
near-IR fluxes are somewhat lower than what is expected 
from the simple extrapolation of the power-law component 
seen in the X-ray band, suggesting that the jet synchrotron 
emission itself is also unlikely to be a main contributor 
to the X-ray flux. These results would justify the 
assumption in our X-ray spectral modeling, that the 
X-ray photons were predominantly produced by 
Comptonization of the disk emission.

\subsection{Implications in the Weak IR and Optical Activity Before the Outburst}
We found weak optical and IR emission of \srcname 
before the start of the 2018 outburst, using the archival 
PTF, WISE, and NEOWISE data. The source 
exhibited $R$-band flux variation by 0.4 mag 
(by 0.4 in the flux unit) on a timescale of $\sim$day, 
and mid-IR variations by $\sim$1 mag (by 2.5) on a 
timescale of $\lesssim$ several days and a few years.
Assuming a distance of 3 kpc, the averaged PTF apparent 
magnitude in the R-band, $\sim 18.5$ mag, is converted to 
an absolute magnitude of $\sim$6 mag. This magnitude 
corresponds to a K-type companion star, if it is a main 
sequence star and dominated the optical flux in the 
PTF observations. 
Because of the significant variations, however, we suggest
that a large fraction of the optical and IR fluxes 
did not originate in the blackbody emission from 
the companion star, but maybe in the outer disk 
and/or jet emission.

\section{Summary and Conclusion}

We have studied the new BHB candidate \srcname
utilizing X-ray data from {\maxi} and {\swift} 
and optical/near-IR data taken in the OISTER
collaboration. What we have found can be summarized as follows.

\begin{enumerate}
\item The source stayed in the low/hard state for $\sim$3 month, 
since its discovery in 2018 March until the start of the second 
brightening in the middle of June. 

\item The X-ray spectrum in that period was successfully described 
with the Comptonization of the disk emission in the hot inner flow or 
corona with electron temperature of $\sim$50 keV.

\item The source showed X-ray short-term variation on timescales of 
seconds, which is likely associated with a change in the 
properties of the Comptonized cloud. 

\item The source exhibited weak activity in the optical 
and near-IR bands before the 2018 outburst, when the source was not 
detected in X-rays. 

\item In the outburst, its optical and near-IR fluxes 
were correlated with the X-ray flux. By modeling the multi-wavelength 
SED at the X-ray flux peak in the end of March, the optical and near-IR 
fluxes were found likely to be contributed by the jet synchrotron emission.
\end{enumerate}

\acknowledgments

This research has made use of {\maxi} data provided by RIKEN, JAXA 
and the {\maxi} team. 
Part of this work was financially supported by Grants-in-Aid 
for Scientific Research 16K17672 (MS), 17H06362 (YU, HN, NK), 
16K05301 (HN), and 
JP16J05742 (YT) from the Ministry of Education, Culture, Sports, 
Science and Technology (MEXT) of Japan. 
This work was supported by Optical and Near-Infrared Astronomy 
Inter-University Cooperation Program and the joint research program 
of the Institute for Cosmic Ray Research (ICRR), 
and by JSPS and NSF under the JSPS-NSF Partnerships for International 
Research and Education (PIRE; RI) and Academy for Global Leadership (AGL) 
of Tokyo Institute of Technology (YT).

\vspace{5mm}
\facilities{\maxi(GSC), \swift(XRT and BAT)}

\end{document}